\documentclass{PoS}

\def\gsim{\mathrel{\rlap{\lower4pt\hbox{\hskip1pt$\sim$}}
    \raise1pt\hbox{$>$}}}         
\def\lsim{\mathrel{\rlap{\lower4pt\hbox{\hskip1pt$\sim$}}
    \raise1pt\hbox{$<$}}}         

\usepackage{amsmath}
\usepackage{amssymb}
\usepackage{graphicx,parskip,caption,subcaption,multirow}

\title{The small-$x$ gluon from forward charm production: implications for a 100 TeV proton collider}

\ShortTitle{The small-$x$ gluon from LHCb charm data}

\author{Rhorry Gauld\\
  ETH Zurich, Institut fur theoretische Physik,
  Wolfgang-Paulistr. 27, 8093, Zurich, Switzerland.\\
        E-mail: \email{rgauld@phys.ethz.ch}}

\author{\speaker{Juan Rojo}\\
        Department of Physics and Astronomy,
VU University, De Boelelaan 1081,
1081 HV Amsterdam,\\ 
and Nikhef Theory Group,
Science Park 105, 1098 XG Amsterdam,
The Netherlands. \\
E-mail: \email{j.rojo@vu.nl}}

\author{Emma Slade\\
        Rudolf Peierls Centre for Theoretical Physics, 1 Keble Road,\\
University of Oxford, OX1 3NP Oxford, United Kingdom.\\
E-mail: \email{emma.slade@physics.ox.ac.uk}}

\abstract{
  We review the constraints on the small-$x$ gluon PDF that
  can be derived by exploiting the
  forward $D$ meson production data from the LHCb experiment
  at $\sqrt{s}=5,7$ and 13 TeV.
  We then discuss the phenomenological implications of the resulting improved
  small-$x$ gluon for ultra-high energy astrophysics, in particular
  neutrino telescopes, as well as for the proposed Future Circular
  Collider (FCC) with $\sqrt{s}=100$ TeV.
    We illustrate how at the FCC even electroweak scale cross-sections can become sensitive
    to the small-$x$ region of the quark and gluon PDFs, and then demonstrate how
    the addition of the LHCb heavy meson production
    measurements leads to a reduction
    of PDF uncertainties for various benchmark cross-sections.
}

  \FullConference{XXV International Workshop on Deep-Inelastic Scattering and Related Subjects\\
		3-7 April 2017\\
		University of Birmingham, UK}

\begin{document}

\paragraph{The small-$x$ gluon from LHCb charm data}
The small-$x$ gluon is one of the worse known parton distributions
functions (PDFs) of the proton due to the lack of direct experimental
information~\cite{Butterworth:2015oua,Rojo:2015acz}.
In global PDF analyses, the small-$x$ region is constrained only
by the inclusive and charm HERA structure
function data, whose coverage is limited to $x\gsim 3\cdot 10^{-5}$ for
$Q^2 \gsim 2$ GeV$^2$.
Recently, it has been demonstrated~\cite{Gauld:2015yia,Zenaiev:2015rfa,Cacciari:2015fta}
that it is possible to constrain this small-$x$ region of $g(x,Q)$ by means of
inclusive $D$ and $B$ meson forward production measurements
from the LHCb experiment.
This sensitivity arises because heavy meson production is driven at the
LHC by the $gg$ luminosity, and the unique forward kinematic coverage of LHCb
allows to cover the very small-$x$ region.
A recent combined analysis~\cite{Gauld:2016kpd}
of the LHCb measurements of $D$ meson production
at $\sqrt{s}=5,7$ and 13 TeV
showed that it is possible
to constrain the gluon PDF reasonably well down to $x \simeq 10^{-6}$,
well below the reach of the HERA data.

The differential cross-sections for $D$ meson production are only known
at NLO and are
affected by rather large theory uncertainties, in particular from
scale variations.
Therefore, the inclusion of the LHCb $D$ measurements in the global
PDF fit requires to introduce normalized cross-sections,
\begin{equation}
N_X ^{ij} = \frac{\mathrm{d}^2 \sigma(\text{X TeV})}{\mathrm{d}y_i^D \mathrm{d}(p_T^D)_j} \bigg/ \frac{\mathrm{d}^2 \sigma(\text{X TeV})}{\mathrm{d}y_{\text{ref}}^D \, \mathrm{d}(p_T^D)_j}\, , \qquad
R_{13/X} ^{ij} = \frac{\mathrm{d}^2 \sigma(\text{13 TeV})}{\mathrm{d}y_i^D \mathrm{d}(p_T^D)_j} \bigg/ \frac{\mathrm{d}^2 \sigma(\text{X TeV})}{\mathrm{d}y_{i}^D \mathrm{d}(p_T^D)_j} \,,
\end{equation}
where $X$ represents a center-of-mass energy and $y_{\rm ref}^D$ stands
for a fixed reference
rapidity bin.
In the case of $N_X^{ij}$, a judicious choice of $y_{\rm ref}^D$ ensures that
most theoretical uncertainties are reduced without losing sensitivity
to the small-$x$ gluon PDF.

\begin{figure}[t]
\centering
  \includegraphics[width=.99\linewidth]{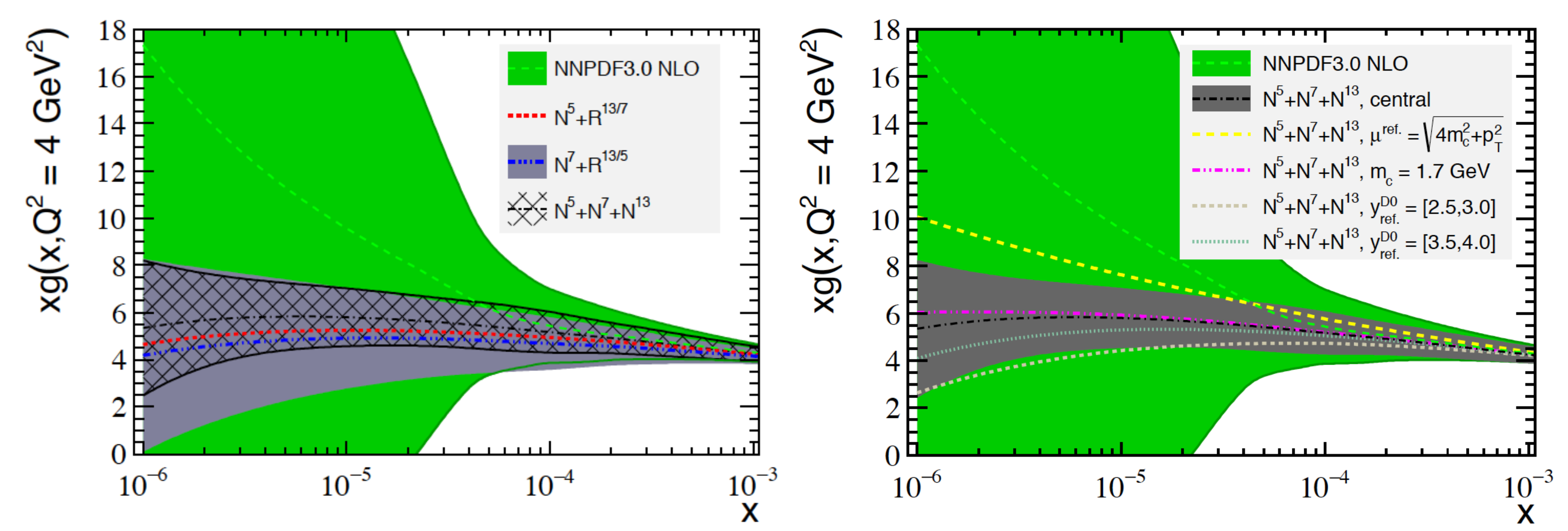}
  \caption{\small Left plot: the small-$x$ gluon PDF in NNPDF3.0 compared with the results
    when various combinations of LHCb $D$ meson production data are included
    in the fit.
    Right plot: the impact of variations of the input theoretical settings
    in the $N^5+N^7+N^{13}$ NNPDF3.0+LHCb fit.
  }
\label{fig:nnpdf30lhc}
\end{figure}

In Fig.~\ref{fig:nnpdf30lhc} we show the small-$x$ gluon
from NNPDF3.0~\cite{Ball:2014uwa} compared with the
the results
    when various combinations of LHCb $D$ meson production data are included
    in the fit~\cite{Gauld:2016kpd}, in particular the
    $N^7+R^{13/5}$ and the $N^5+N^7+N^{13}$ combinations that exhibit
    similar constraining power.
    We observe how at small-$x$ the gluon PDF uncertainties can be reduced
    by up to an order of magnitude as compared to the baseline NNPDF3.0
    results.
    In the same figure we also show
    the impact of the variations of the input theoretical settings
    in the $N^5+N^7+N^{13}$ NNPDF3.0+LHCb fit, illustrating than
    in all cases the resulting shifts in the small-$x$ gluon
    are subdominant or of similar size as the
    resulting PDF uncertainties,
    and thus demonstrating the robustness
    of the analysis.

\paragraph{Implications for high-energy neutrino telescopes}

\begin{figure}[t]
\centering
  \includegraphics[width=.99\linewidth]{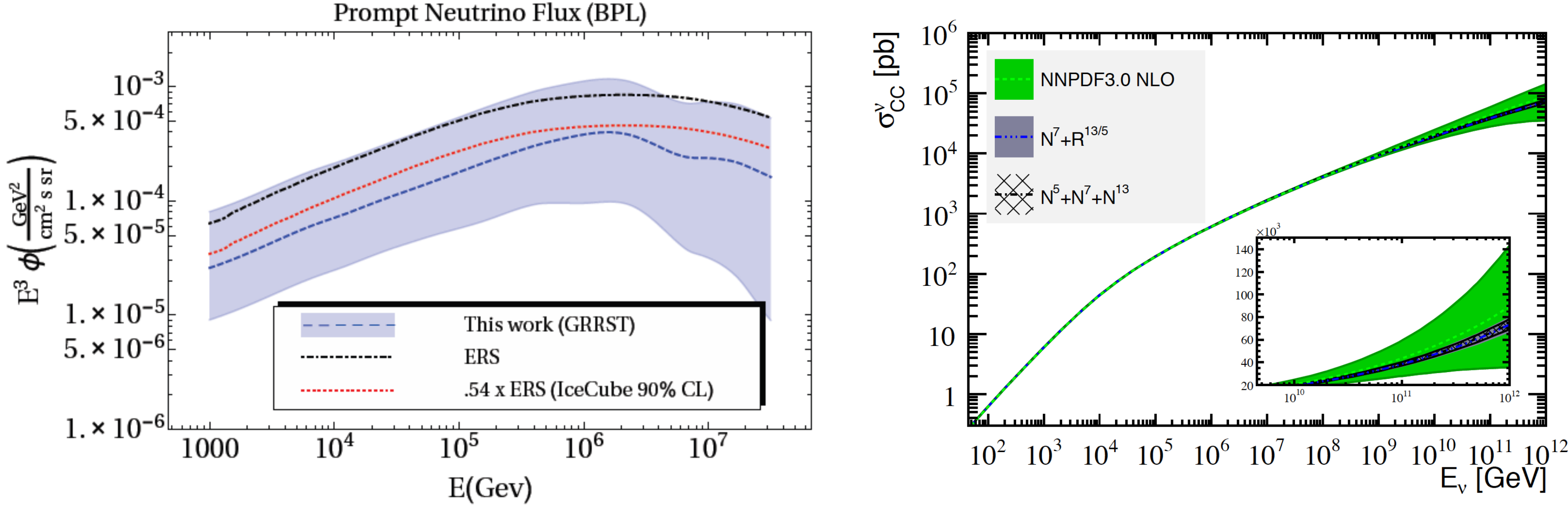}
  \caption{\small Left plot: the GRRST prompt neutrino flux computed 
    using the $N^7$ NNPDF3.0+LHCb set with the corresponding
    theory uncertainty band.
    We also show the results of a recent upper bound from the IceCube
    data analysis.
    Right plot: the charged current UHE neutrino-nucleus cross-section
    computed with the NNPDF3.0+LHCb sets, compared with the NNPDF3.0 baseline.
  }
\label{fig:pheno-astro}
\end{figure}

In order to illustrate the implications of the NNPDF3.0+LHCb sets
for ultra high energy (UHE)
astroparticle physics, in Fig.~\ref{fig:pheno-astro} we show 
the prompt neutrino flux from Ref.~\cite{Gauld:2015kvh}
computed in
using the NNPDF3.0+LHCb results of
Ref.~\cite{Gauld:2015kvh} (GRRST) together
with the corresponding theory uncertainty band.
This prompt flux arises from the decays of $D$ and $B$ mesons produced
in cosmic ray collisions in the atmosphere, and is the dominant background
for astrophysical UHE neutrinos.
We also show the results of the upper bounds from a recent IceCube
data analysis.
The IceCube bounds are are slightly below the GRRST prediction,
 suggesting that a first direct detection of the prompt
neutrino flux could be within reach.

In Fig.~\ref{fig:pheno-astro} we also show
     the charged-current (CC) UHE neutrino-nucleus cross-section
     computed with the NNPDF3.0+LHCb sets.
     We find that as a consequence of the reduction
     of the small-$x$ gluon PDF uncertainties, now few-percent
     theory errors can be achieved up to the most extreme
    neutrino energies, $E_{\nu}=10^{12}$ GeV.
    Having a robust QCD baseline for UHE neutrino-nucleus
    cross-sections paves the way for novel studies, both in the SM
    and beyond it, based on upcoming data from neutrino telescopes at these
    extreme energies.
    One  example 
    would be testing for departures of the linear DGLAP
    framework by searching for non-linear effects small-$x$ (BFKL) resummation.

\paragraph{Benchmark cross-sections at 100 TeV}
At the proposed Future Circular Collider (FCC),
operating with a centre of mass energy of $\sqrt{s} = 100$ TeV, many
processes will be sensitive to the
PDFs down to the very low-$x$ region, $x \lesssim 10^{-5}$, even relatively
high-scale processes such as 
inclusive $W$ or $Z$ production~\cite{Mangano:2016jyj,Rojo:2016kwu}.
The accurate determination of the PDFs in this regime is therefore
an important ingredient of the FCC physics program.
Here we explore the impact that the improved small-$x$ PDFs constrained by the LHCb $D$ meson
data, derived in~\cite{Gauld:2016kpd},
have on a number of important cross-sections for proton-proton
collisions at $\sqrt{s}=100$ TeV.

First of all, we have computed fiducial cross sections for $W$ and $Z$ production
using MCFM~\cite{Boughezal:2016wmq}.
We have produced NLO results
for the baseline NNPDF3.0 and for two
of the NNPDF3.0+LHCb sets, namely those based on
the $N_7$+$R_{13/5}$
and $N_5$+$N_7$+$N_{13}$ combinations.
We have considered
three different possibilities for the acceptance cuts on the kinematics
of the final-state leptons: i) no cuts,
ii) ``LHC cuts'' of $p_T ^l \geq 20$ GeV and $|y_l| \leq 2.5$ and 
iii) ``FCC cuts'' cuts of $p_T ^l \geq 20$ GeV and $|y_l| \leq 5$.
The latter cuts are motivated by the expectation that the FCC should
have improved coverage of the forward region as compared to the LHC.
For $Z$ production, the dilepton invariant mass is restricted to the region
$66 \ge m_{ll} \ge 116$ GeV.

\begin{figure}[t]
\centering
  \includegraphics[width=.49\linewidth]{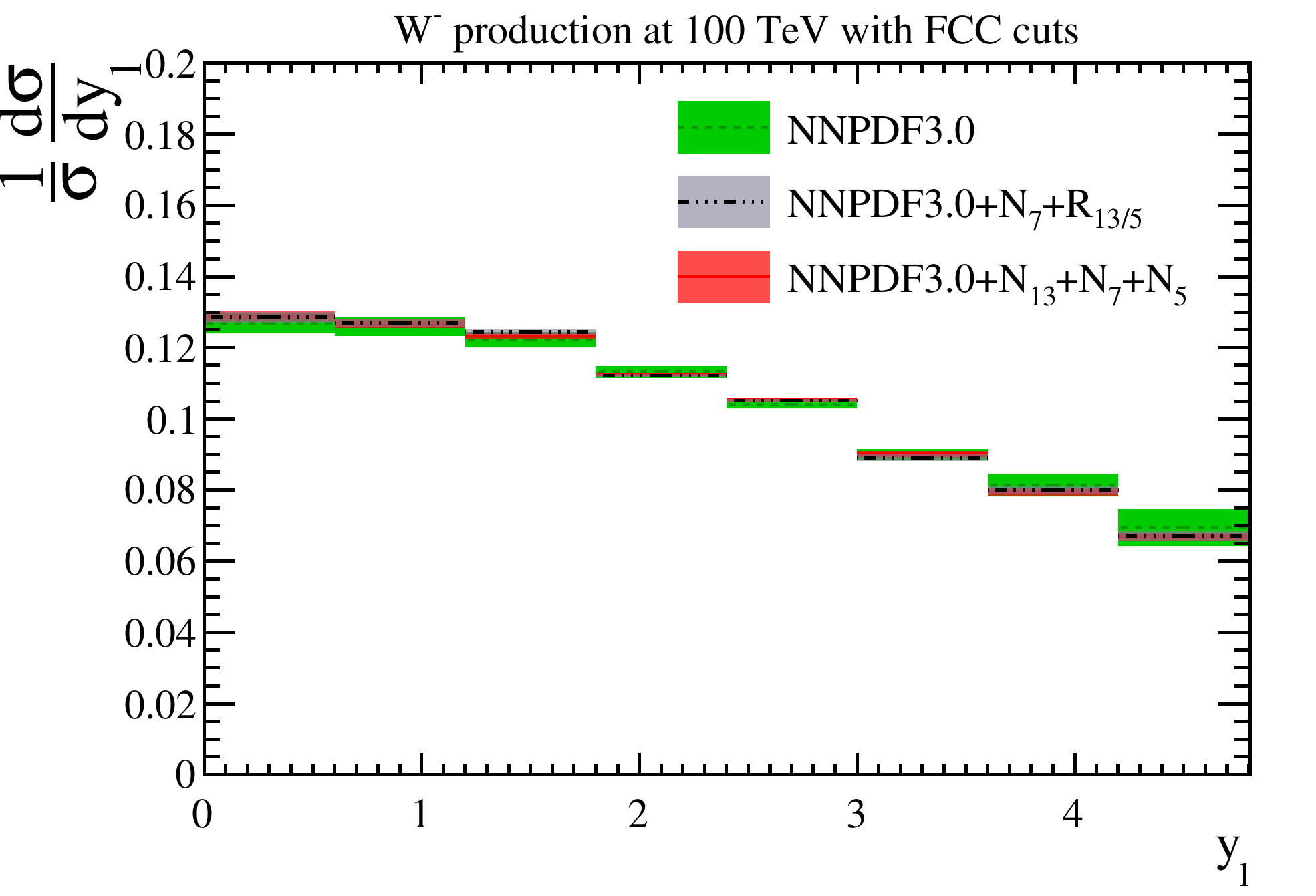}
  \includegraphics[width=.49\linewidth]{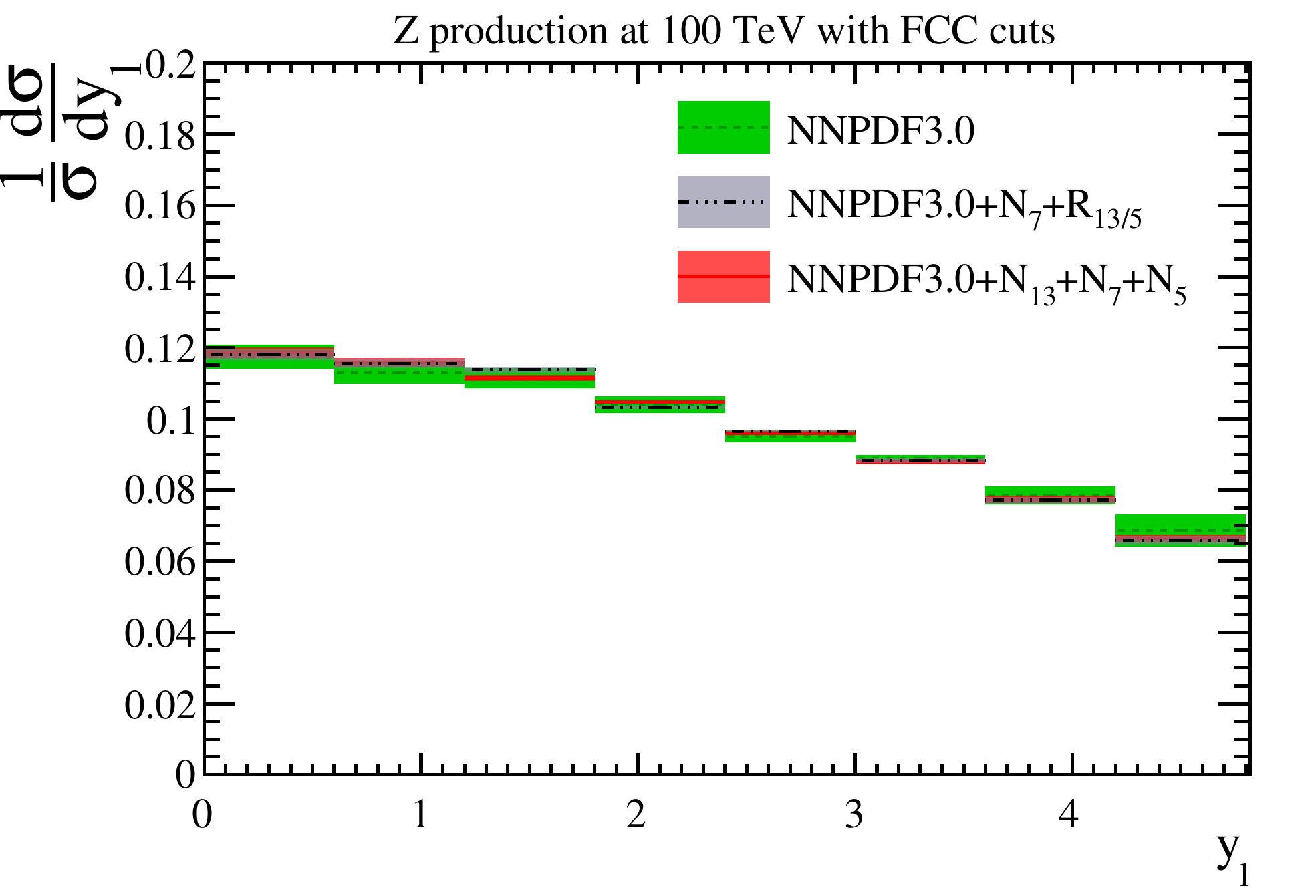}
  \caption{\small Normalised charged lepton rapidity distributions from
    $W^-$ (left) and $Z$ (right) boson production at $\sqrt{s}$ = 100 TeV.
    We show the predictions from NNPDF3.0 and for the two NNPDF3.0+LHCb sets,
    together with the
    corresponding PDF uncertainties.
  }
\label{fig:wma}
\end{figure}

In Fig.~\ref{fig:wma} we show the normalised rapidity distributions of the charged leptons from
$W^-$ and $Z$ decays corresponding to the ``FCC cuts'' scenario, namely where the inclusive cross-section receives contributions from lepton
rapidities as large as $|y_l|=5$.
Taking into account the LO kinematics,
$x_{1,2}=(M_V/\sqrt{s})e^{\pm y_V}$,
where $y_V$ is the gauge boson rapidity, one sees that
that PDFs down to $x\simeq 10^{-5} $ are being probed.
As expected, we observe a
reduction in the corresponding PDF uncertainties specially
at high rapidities, which is where the
sensitivity to the small-$x$ gluon comes from,
once the NNPDF3.0+LHCb sets are used as compared to the baseline
NNPDF3.0 results.

This reduction of the PDF uncertainties at large lepton
rapidities has direct consequences for the
inclusive $W$ and $Z$ cross-sections integrated over rapidity.
In Table~\ref{table:14tevand100TeV} we list the
cross sections times
branching ratios $\sigma(V){\rm BR}(V\to l_1l_2)$ for weak gauge boson production at 14 TeV and 100 TeV for different sets of kinematic cuts.
  We indicate both the central value (in nb) and the percentage
  PDF uncertainty for the NNPDF3.0 and NNPDF3.0+LHCb NLO sets,
  the latter based on the $N^5+N^7+N^{13}$ combination.
  We see that while at 14 TeV the PDF uncertainties are of the same
  size with and without cuts, at 100 TeV this is not the case:
  without cuts, the inclusive $W,Z$ cross-sections exhibit
  PDF uncertainties as large as 7\%, and even with the ``FCC cuts'' they
  can be up to a factor 2 larger than those corresponding to the
  ``LHC cuts'', reflecting the sensitivity to the small-$x$ region.
  For the NNPDF3.0+LHCb set instead, the PDF uncertainties
  become essentially independent of the choice of kinematical cuts,
  highlighting the stabilization of the small-$x$ region.

\begin{table}[h]
  \small \centering
\begin{tabular}{|l | c || c | c ||  c | c | c |}
\hline
& & \multicolumn{2}{c|}{14 TeV} & \multicolumn{3}{c|}{100 TeV} \\ \hline
& & No cuts & LHC cuts & No cuts & LHC cuts & FCC cuts \\ \hline
\hline
\multirow{3}{*}{NNPDF3.0}& $W^+$  & 11.8 (1.9\%) & 6.4 (2.0\%) & 73.5 (7.0\%) & 27.8 (2.9\%)  & 52.8 (4.9\%) \\
& $W^-$ & 8.8 (1.8\%)& 4.7 (1.4\%) & 61.9 (5.5\%) & 26.0 (3.0\%) & 44.1 (3.6\%) \\
& $Z$ & 2.0 (1.7\%) & 1.5 (1.8\%) & 14.1 (5.1\%) & 7.9 (3.2\%) & 12.5 (4.1\%)
\\
\hline
  \hline  
\multirow{3}{*}{NNPDF3.0+LHCb}&  $W^+$  & 12.2 (1.6\%) & 6.6 (1.7\%) & 73.4 (3.0\%) & 29.0 (2.7\%) &
  53.5 (2.8\%) \\
&  $W^-$ & 9.1 (1.6\%) & 4.9 (1.7\%) & 62.3 (2.9\%) & 27.2 (2.8\%)
  & 45.2 (2.8\%) \\
& $Z$ & 2.1 (1.6\%) &  1.5 (1.7\%) & 14.3 (2.8\%) & 8.3 (2.9\%) & 12.8 (2.8\%) 
\\ \hline
\end{tabular}
\caption{\small The cross sections times
  branching ratios $\sigma(V){\rm BR}(V\to l_1l_2)$ for weak gauge boson
  production
  at 14 TeV and 100 TeV for different sets of kinematic cuts, described
  in the text.
  We indicate both the central value (in nb) and the percentage
  PDF uncertainty for the NNPDF3.0 and NNPDF3.0+LHCb NLO sets,
  the latter based on the $N^5+N^7+N^{13}$ combination.
  \label{table:14tevand100TeV}
  }
\end{table}

In addition to inclusive $W/Z$ production, we have calculated other benchmark cross-sections
at 100 TeV.
To begin with, we have calculated inclusive
prompt photon production at NLO with MCFM, where the following fiducial cuts have
been imposed: $|y_\gamma| < 5$ and $p_T ^\gamma > 20$ GeV.
Then we have also computed the cross-sections for Drell-Yan $\gamma^*/Z$ production
at low invariant masses, where we require $p_T^l \ge 20$ GeV, $|y_l| \leq 5$,
and $20 < m_{ll} < 30$ GeV for the final-state leptons.
Both process are characterized by some sensitivity to the small-$x$ region.
Moreover, we have also computed heavy quark pair production both for charm
and for bottom quarks.
In this case we have assumed that the FCC would be equipped with the forward
detector, analogous to the LHCb experiment, that can instrument the region
$2.5 < y_Q < 6.5 $ for the heavy quark rapidity.
Heavy quark pair production cross-sections have been computed at NLO using \texttt{MadGraph\_aMC@NLO}~\cite{Alwall:2014hca}
without imposing any cut on the heavy quark $p_T^Q$.

\begin{figure}[t]
  \centering
  \includegraphics[width=.49\linewidth]{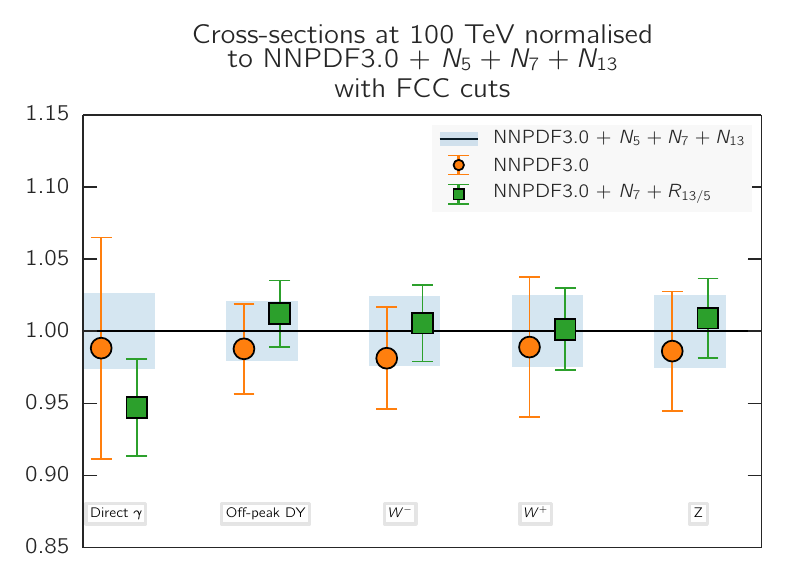}
  \includegraphics[width=.49\linewidth]{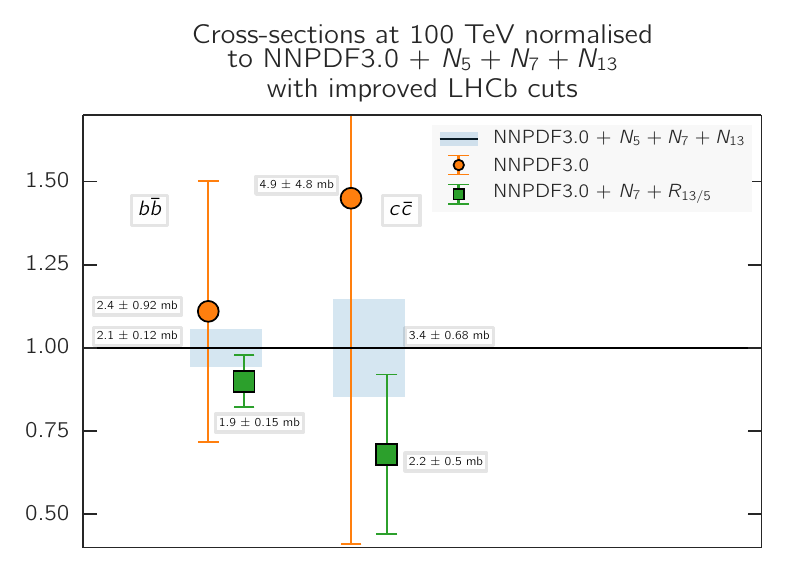}
  \caption{\small Left plot: inclusive cross sections at NLO with ``FCC cuts'' (see text)
    computed with NNPDF3.0 and two of the NNPDF3.0+LHCb sets, normalized
    to the central value of the $N_5+N_7+N_{13}$ fit.
    The error bands includes only the PDF uncertainties.
    Right plot: same for charm and bottom pair production,
    in the fiducial region defined by $2.5 < y_Q < 6.5$.
    }
\label{fig:tot1}
\end{figure}
 
The overview of the results is presented in Fig.~\ref{fig:tot1},
where we compare the various NLO cross-sections, with the ``FCC cuts'', computed with
NNPDF3.0 and the two NNPDF3.0+LHCb sets.
In all cases we observe a reduction of the PDF uncertainties as a consequence
of the improved small-$x$ gluon PDF.
For instance, for direct photon production the PDF uncertainties become smaller
by more than a factor of two.
Unsurprisingly, the effect is the largest for forward heavy quark production,
specially in the charm quark case.
Using NNPDF3.0, the PDF uncertainties in the fiducial 100 TeV cross-section are around
100\%, which are reduced to around 20\% once the NNPDF3.0+LHCb sets are used.
Fig.~\ref{fig:tot1} thus illustrates how the exploitation of available and future
LHC data can improve the prospects of a precision physics program at a future
100 TeV hadron collider.
Another example would be the use of top-quark pair differential distributions
in the NNLO global fit to constrain the gluon PDF at large $x$~\cite{Czakon:2016olj},
the region relevant for new heavy BSM resonances produced  in the gluon-gluon channel.

\providecommand{\href}[2]{#2}\begingroup\raggedright\endgroup


\begin{thebibliography}{10}

\bibitem{Butterworth:2015oua}
J.~Butterworth et~al., {\it {PDF4LHC recommendations for LHC Run II}},  {\em J.
  Phys.} {\bf G43} (2016) 023001, [\href{http://arxiv.org/abs/1510.03865}{{\tt
  arXiv:1510.03865}}].

\bibitem{Rojo:2015acz}
J.~Rojo et~al., {\it {The PDF4LHC report on PDFs and LHC data: Results from Run
  I and preparation for Run II}},  {\em J. Phys.} {\bf G42} (2015) 103103,
  [\href{http://arxiv.org/abs/1507.00556}{{\tt arXiv:1507.00556}}].

\bibitem{Gauld:2015yia}
R.~Gauld, J.~Rojo, L.~Rottoli, and J.~Talbert, {\it {Charm production in the
  forward region: constraints on the small-x gluon and backgrounds for neutrino
  astronomy}},  {\em JHEP} {\bf 11} (2015) 009,
  [\href{http://arxiv.org/abs/1506.08025}{{\tt arXiv:1506.08025}}].

\bibitem{Zenaiev:2015rfa}
{\bf PROSA} Collaboration, O.~Zenaiev et~al., {\it {Impact of heavy-flavour
  production cross sections measured by the LHCb experiment on parton
  distribution functions at low x}},  {\em Eur. Phys. J.} {\bf C75} (2015),
  no.~8 396, [\href{http://arxiv.org/abs/1503.04581}{{\tt arXiv:1503.04581}}].

\bibitem{Cacciari:2015fta}
M.~Cacciari, M.~L. Mangano, and P.~Nason, {\it {Gluon PDF constraints from the
  ratio of forward heavy-quark production at the LHC at $\sqrt{S}=7$ and 13
  TeV}},  {\em Eur. Phys. J.} {\bf C75} (2015), no.~12 610,
  [\href{http://arxiv.org/abs/1507.06197}{{\tt arXiv:1507.06197}}].

\bibitem{Gauld:2016kpd}
R.~Gauld and J.~Rojo, {\it {Precision determination of the small-$x$ gluon from
  charm production at LHCb}},  {\em Phys. Rev. Lett.} {\bf 118} (2017), no.~7
  072001, [\href{http://arxiv.org/abs/1610.09373}{{\tt arXiv:1610.09373}}].

\bibitem{Ball:2014uwa}
{\bf NNPDF} Collaboration, R.~D. Ball et~al., {\it {Parton distributions for
  the LHC Run II}},  {\em JHEP} {\bf 04} (2015) 040,
  [\href{http://arxiv.org/abs/1410.8849}{{\tt arXiv:1410.8849}}].

\bibitem{Gauld:2015kvh}
R.~Gauld, J.~Rojo, L.~Rottoli, S.~Sarkar, and J.~Talbert, {\it {The prompt
  atmospheric neutrino flux in the light of LHCb}},  {\em JHEP} {\bf 02} (2016)
  130, [\href{http://arxiv.org/abs/1511.06346}{{\tt arXiv:1511.06346}}].

\bibitem{Mangano:2016jyj}
M.~L. Mangano et~al., {\it {Physics at a 100 TeV pp collider: Standard Model
  processes}},  \href{http://arxiv.org/abs/1607.01831}{{\tt arXiv:1607.01831}}.

\bibitem{Rojo:2016kwu}
J.~Rojo, {\it {Parton Distributions at a 100 TeV Hadron Collider}},  {\em PoS}
  {\bf DIS2016} (2016) 275, [\href{http://arxiv.org/abs/1605.08302}{{\tt
  arXiv:1605.08302}}].

\bibitem{Boughezal:2016wmq}
R.~Boughezal, J.~M. Campbell, R.~K. Ellis, C.~Focke, W.~Giele, X.~Liu,
  F.~Petriello, and C.~Williams, {\it {Color singlet production at NNLO in
  MCFM}},  {\em Eur. Phys. J.} {\bf C77} (2017), no.~1 7,
  [\href{http://arxiv.org/abs/1605.08011}{{\tt arXiv:1605.08011}}].

\bibitem{Alwall:2014hca}
J.~Alwall, R.~Frederix, S.~Frixione, V.~Hirschi, F.~Maltoni, et~al., {\it {The
  automated computation of tree-level and next-to-leading order differential
  cross sections, and their matching to parton shower simulations}},  {\em
  JHEP} {\bf 1407} (2014) 079, [\href{http://arxiv.org/abs/1405.0301}{{\tt
  arXiv:1405.0301}}].

\bibitem{Czakon:2016olj}
M.~Czakon, N.~P. Hartland, A.~Mitov, E.~R. Nocera, and J.~Rojo, {\it {Pinning
  down the large-x gluon with NNLO top-quark pair differential distributions}},
   {\em JHEP} {\bf 04} (2017) 044, [\href{http://arxiv.org/abs/1611.08609}{{\tt
  arXiv:1611.08609}}].

\end{thebibliography}

\end{document}